\documentclass{scrartcl}
\usepackage[utf8]{inputenc}
\usepackage{authblk}
\usepackage[numbers,sort&compress]{natbib}
\usepackage{hyperref}
  \hypersetup{colorlinks,linkcolor=[rgb]{0,0,0.5},citecolor=[rgb]{0,0,0.5},urlcolor=[rgb]{0,0,0.5},breaklinks}
\usepackage{doi}
\usepackage{amsmath}
\usepackage{graphicx}
\usepackage{fancyhdr}
\usepackage{multirow}
\usepackage{tikz} 
\usepackage{soul}

\definecolor{lime}{HTML}{A6CE39}
\DeclareRobustCommand{\orcidicon}{
	\hspace{-2.7mm}
	\begin{tikzpicture}
	\draw[lime, fill=lime] (0,0) 
	circle [radius=0.16] 
	node[white] {{\fontfamily{qag}\selectfont \tiny ID}};
	\draw[white, fill=white] (-0.0625,0.095) 
	circle [radius=0.007];
	\end{tikzpicture}
	\hspace{-2.7mm}
}
\newcommand{\orcid}[1]{\href{https://orcid.org/#1}{\orcidicon}}

\title{Enhanced Preconditioner for JOREK MHD Solver}

\author[1,2]{I Holod\orcid{0000-0002-2078-9489}}
\author[2]{M Hoelzl\orcid{0000-0001-7921-9176}}
\author[2]{P S Verma}
\author[3,4]{GTA Huijsmans}
\author[5,6]{R Nies\orcid{0000-0002-9508-1223}}
\author[7]{JOREK Team}

\affil[1]{Max Planck Computing and Data Facility (MPCDF)}
\affil[2]{Max Planck Institute for Plasma Physics, Boltzmannstr. 2, 85748 Garching b. M., Germany}
\affil[3]{CEA, IRFM, 13108 Saint-Paul-Lez-Durance, France}
\affil[4]{Eindhoven University of Technology, P.O. Box 513, 5600 MB Eindhoven, The Netherlands}
\affil[5]{Department of Astrophysical Sciences, Princeton University, Princeton, NJ, 08543, USA}
\affil[6]{Princeton Plasma Physics Laboratory, Princeton, NJ, 08540, USA}
\affil[7]{See the author list of [M Hoelzl, G T A Huijsmans, S J P Pamela, M Becoulet, E Nardon, F J Artola, B Nkonga et al, Nuclear Fusion (submitted); pre-print at \url{https://arxiv.org/abs/2011.09120}]}

\makeatletter
\renewcommand\maketitle{
   \begin{center}
     {\huge\sffamily\bfseries\@title\par\vspace{0.3em}}
     {\scshape\large\@author}
   \end{center}
}
\makeatother

\begin{document}

\maketitle

{\small
\tableofcontents
}

\section*{Abstract}

The JOREK extended magneto-hydrodynamic (MHD) code is a widely used simulation tool for studying the non-linear dynamics of large-scale instabilities in divertor tokamak plasmas. Due to the large scale-separation intrinsic to these phenomena both in space and time, the computational costs for simulations in realistic geometry and with realistic parameters can be very high, motivating the investment of considerable effort for optimization. The code is usually run with a fully implicit time integration allowing to use large time steps independent of a CFL criterion. This is particularly important due to the strong scale separation between transport processes and slowly growing resistive modes in contrast to fast time scales associated with MHD waves and fast parallel heat transport. For solving the resulting large sparse-matrix system iteratively in each time step, a physics-based preconditioner built on the assumption of weak coupling between the toroidal harmonics is applied. The solution for each harmonic matrix is determined independently in this preconditioner using a direct solver. In this article, a set of developments regarding the JOREK solver and preconditioner is described, which lead to overall significant benefits for large production simulations. This comprises in particular enhanced convergence in highly non-linear scenarios and a general reduction of memory consumption and computational costs. The developments include the implementation of a complex solver interface for the preconditioner. The most significant development presented consists in a generalization of the physics based preconditioner to ``mode groups'', which allows to account for the dominant interactions between toroidal Fourier modes in highly non-linear simulations. At the cost of a moderate increase of memory consumption, the technique can strongly enhance convergence in suitable cases allowing to use significantly larger time steps. For all developments, benchmarks based on typical simulation cases demonstrate the resulting improvements.

\section{Introduction}\label{:intro}

Non-linear extended MHD allows to describe a broad variety of different large-scale instabilities in magnetically confined fusion plasmas accurately and with lower computational costs than (gyro)kinetic approaches. Numerically, however, there are significant challenges to overcome: large spatial and temporal scale separations need to be bridged. Highly anisotropic heat transport, fast plasma flows and a large scale separation in time render it virtually impossible to simulate realistic parameters without an implicit time integrator.

The following Sections~\ref{:intro:jorek} to~\ref{:intro:setup} provide background on the JOREK code and the sparse matrix solvers, explain the numerical stages executed during a time step, and describe the setup used for the numerical tests. Section~\ref{:intro:outline} contains the outline of the main part of the article, which describes the developments implemented before they are tested via a series of benchmarks.

\subsection{JOREK overview}\label{:intro:jorek}

JOREK~\cite{Huysmans2007,Hoelzl2020B} is a state-of-the-art code for the simulation of large-scale instabilities in tokamaks (magnetic confinement fusion devices). It solves extended  reduced~\cite{Orain2013} or full~\cite{Pamela2020} MHD equations using a 2D iso-parametric $G^1$ continuous finite element discretization ~\cite{Czarny2008}, which was recently extended to arbitrary order \cite{Pamela2021}, combined with a toroidal Fourier decomposition. Main applications are simulations addressing the dynamics of major disruptions including vertical displacement events and runaway electrons as well as edge localized modes. The goal is to advance the physics understanding for these potentially very harmful phenomena and to develop reliable control schemes applicable in present and future fusion devices.

A free boundary and resistive wall extension exists~\cite{Hoelzl2012B} and a framework for kinetic effects is available~\cite{vanVugtPhD}.
The time integration is fully implicit such that a very large -- and, due to the properties of the equations, badly conditioned -- sparse matrix system needs to be solved in every time step. The present article describes improvements of the computational efficiency of the code by a variety of different approaches enhancing the solver performance. These methods may also be beneficial in other codes.
We describe the key properties of the JOREK solver and the sparse matrix system in the following Subsections~\ref{:intro:jorek:solver} and~\ref{:intro:jorek:matrix}.

\subsubsection{JOREK Solver}\label{:intro:jorek:solver}
Starting with the general form of MHD equations
\begin{equation*}
\frac{\partial\mathbf{A}(\mathbf{u})}{\partial t}=\mathbf{B}(\mathbf{u},t),
\end{equation*}
we get the following linearized with respect to $\delta\mathbf{u}$ time discretization
\begin{equation}\label{eq:CN}
\left[\left(1+\zeta\right)\left(\frac{\partial\mathbf{A}}{\partial\mathbf{u}}\right)^n - 
\Delta t\theta\left(\frac{\partial\mathbf{B}}{\partial\mathbf{u}}\right)^n\right]\delta\mathbf{u}^n = \Delta t\mathbf{B}^n + \zeta\left(\frac{\partial\mathbf{A}}{\partial\mathbf{u}}\right)^{n-1}\delta\mathbf{u}^{n-1}.
\end{equation}
Here, $n$ represents the time step index, and $\delta\mathbf{u}=\delta\mathbf{u}^{n+1}-\delta\mathbf{u}^n$ where $\mathbf{u}$ is a vector of all unknowns.  Setting $\theta=1/2$ and $\zeta=0$, Eq.~\ref{eq:CN} recovers the Crank-Nicolson scheme, while alternatively setting $\theta=1$ and $\zeta=1/2$ the BDF2 (Gears) scheme is recovered.

After spatial discretization, Eq.~\ref{eq:CN} forms a large linear algebraic system of equations 
\begin{equation}\label{eq:AX}
A\,x=b
\end{equation}
where A is a sparse matrix. Two main options are available for that: when the problem size is small, in particular for axisymmetric cases, a direct sparse matrix solver is applied to the complete system. The second option is applied for the large problem sizes typically encountered in real-life applications: the system is solved iteratively with the generalized minimal residual (GMRES) method \cite{saad86}. 

Due to the stiffness of the system, e.g. caused by the highly anisotropic diffusion operator and the presence of large convective terms in MHD equations the matrix $A$ is ill-conditioned, i.e. a small change in the right hand side $b$ leads to a big change of the solution $x$. This causes extremely poor convergence of the GMRES algorithm. To improve the convergence, (left-) preconditioning (PC) is applied \cite{saadbook}, namely the original system (\ref{eq:AX}) is replaced by 
\begin{equation}\label{eq:PC1}
M^{-1}A\,x=M^{-1}b
\end{equation}
where $M$ is a non-singular preconditioner matrix, which should be close to A in some sense, and the solution
\begin{equation}\label{eq:PC2}
z=M^{-1}w
\end{equation}
should be easy to find in comparison to solving the original system. At the first step of the GMRES iteration cycle $w=b$, and later on $w=b-Ax_m$, where $x_m$ is the solution of (\ref{eq:PC1}) at $m$-th iteration. It is noteworthy that the matrix $M$ is only involved in the algorithm for obtaining the solution vectors.

\begin{figure}
\centering
   \includegraphics[height=0.3\textwidth]{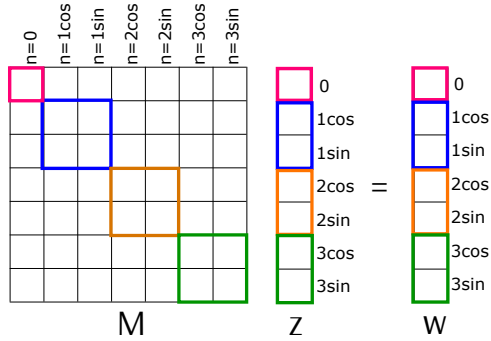}
\caption{The matrix structure is shown for a simple example case with the toroidal modes $n=(0,1,2,3)$. The color blocks outline the parts of the original matrix $A$ used to form the preconditioner solution $z=M^{-1}\,w$. Each block represent individual toroidal Fourier mode.}
\label{fig:modes}
\end{figure}

In the standard JOREK algorithm, the preconditioner matrix $M$ has block-diagonal structure with each block corresponding to the individual toroidal Fourier modes. Physically this corresponds to the approximation of $A$ by omitting mode-coupling terms. A schematic representation for a small case with the toroidal modes $n=(0,1,2,3)$ is shown in Figure~\ref{fig:modes}. The solution vector $z$ is then formed by combining the individual solutions for each diagonal block, which can be found independently by using direct sparse solver libraries. The advantage of this method is it's scalability with respect to the toroidal resolution and robustness: the underlying physics leading to ill-conditioning is addressed by using the direct solver applied to the PC matrices.

The preconditioning matrix $M$ can usually be reused for many time steps thus saving on the expensive LU-factorization of the blocks. In the linear phase of the physical system's evolution, when the mode amplitudes are small and mode coupling is negligible, the preconditioner matrix $M$ is an almost exact approximation of the full matrix $A$ such that the iterative solver usually converges in a single iteration. However, when the problem becomes non-linear such that the coupling between toroidal modes cannot be neglected any more, the preconditioner becomes far less efficient, and consequently the GMRES convergence deteriorates. This poor approximation of $A$ by the preconditioning matrix $M$ is one of the main challenges for the efficiency of the JOREK solver. Besides that, the large memory consumption of the LU decomposed preconditioner matrix blocks and the limited parallel scalability of the direct sparse matrix solvers are additional challenges to be faced. 

In the present work, we address these challenges, leading to a significant step forward in code efficiency. This includes a complex representation of preconditioner block-matrices (Section~\ref{:dev:compl}) and a generalization of the preconditioner allowing to approximate the matrix $A$ more accurately in highly non-linear scenarios (Section~\ref{:dev:modegroups}). The performance of the new developments is evaluated in Section~\ref{:benchmarks}.

\subsubsection{Matrix properties}\label{:intro:jorek:matrix}

Here we describe the sparse matrix properties for a typical large problem size of 30 thousand grid nodes, 8 physical variables, 4 degrees of freedom per grid node, and 41 toroidal harmonics -- this is counting cosine and sine components separately such that this corresponds to toroidal modes $n=(0\dots20)$.

\paragraph{Global matrix}
The total dimension of the matrix corresponding to this problem size is about 40 million. All node variables are typically coupled with each other, and, additionally, with eight neighboring grid nodes\footnote{At special points like the grid axis and the X-point, this may be different. Depending on the boundary conditions, the connectivity can also be different there.}. This leads to about 12 thousand non-zero entries in each matrix row and about 500 billion non-zero entries in the whole matrix, which requires about 4 TB of main memory for storing the double precision floating point numbers. Additional memory is needed for storing the matrix indices, depending on the sparse matrix format used. For the coordinate format using 2 integer indices of 4-byte each, the total memory requirement for the matrix storage is about 8 TB. In case of compressed sparse row (CSR) format, the amount of memory needed to store the location of the entries is reduced by nearly 50\%, so totally a bit more than 6 TB would be required. In JOREK, the matrix is constructed in coordinate format and the preconditioner matrices are stored in CSR format.

As shown above, due to the locality of the 2D Bezier basis functions, the global matrix is sparse with only one out of 3000 matrix entries different from zero in each row or, respectively, column. Since the matrix in our example does not fit into the memory of one typical compute node (100-200 GB), we use domain decomposition to construct matrix in a distributed way. If in this case we use 84 compute nodes with 4 MPI tasks per node (336 MPI tasks in total), each MPI task is responsible for constructing the matrix contributions by about 90 grid elements. This way, roughly 25 GB of storage are needed per MPI task for the distributed global matrix, i.e., about 100 GB per compute node. The matrix construction on each MPI task is OpenMP parallelized. If we assume 8 OpenMP threads per MPI task, each thread is responsible for creating the matrix contributions of about 10 grid elements (called elementary matrices). The elementary matrices are dense and have a dimension of about 5000 (4 nodes times 4 degrees of freedom times 8 variables times 41 Fourier harmonics) requiring about 200 MB of main memory each. The matrix is constructed using fast Fourier transform. When a single toroidal Fourier mode is used (in test simulation), direct integration is applied instead.

\paragraph{Preconditioner matrix}
For the considered example, each block of the preconditioner matrix $P$ has a dimension of about 2 million, and about 600 non-zero entries per matrix row, such that each of the block-matrices has about 1 billion non-zero entries consuming about 12 GB of main memory in CSR format. Each of these blocks corresponds to a particular toroidal mode $n$, and need to be LU decomposed in the preconditioning, see Section~\ref{:intro:jorek:solver}. All blocks together require about 250 GB of memory. The block-matrix corresponding to $n=0$ is smaller by a factor of two in dimension and by a factor of four in the number of non-zero entries (since $n=0$ has only axisymmetric component). During LU-factorization (see Section~\ref{:intro:jorek:solver}), the memory consumption increases significantly due to the fill-in.

\subsection{Phases in a time step}\label{:intro:phases}

Now that the JOREK solver has been introduced, we identify the major numerical stages performed during a single time step. The stages are abbreviated by one letter in the following to simplify notation in figures and tables later on. Note, that the present article concentrates on the phases after the global matrix construction. Here, \textbf{F} and \textbf{I} are usually most critical for the overall performance.
\begin{itemize}
\item \textbf{G: Global matrix construction.} Matrix $A$ is constructed in a distributed way at every time step. This can cost a significant fraction of the overall run time depending on the case.
\item \textbf{P: Construction of PC block-matrices.} So far done by extracting from the global matrix via communication; an alternative option for direct construction was implemented. This part berries a significant computational cost comparable to \textbf{G}, however it is not usually performed at every time step thus not as critical.
\item \textbf{A: Analysis of the PC matrix structure.} Here matrix is analyzed and elimination tree is build. This is only done in the very first time step of a simulation, since the structure is not changing during a simulation, but only the values of the matrix entries. This phase is therefore usually not critical for the overall performance.
\item \textbf{F: LU-factorization of the PC matrices.} Done only in time steps where the preconditioner needs to be updated. Often critical for the overall performance. The relative costs compared to phase~\textbf{I} depend on the application and time step used, but also on the user defined threshold for updating the preconditioning matrix -- the user can influence the frequency of the preconditioner updates to some degree via the choice of the time step and via a threshold.
\item \textbf{I: Iterative solution of the global system} including solver calls for the preconditioner matrix blocks. Done in every time step. Often critical for the overall performance. The relative costs compared to the preconditioner updates of phase~\textbf{F} depend on case and settings (see explanation above).
\end{itemize}

\subsection{Setup for the numerical tests}\label{:intro:setup}

The quantities listed in the following are used in this article to describe the benchmark setups and results. \newline
\begin{itemize}
\item $N_\text{N}$: Total number of compute nodes used in the simulation
\item $N_\text{M}$: Total number of MPI tasks used in the simulation
\item $M_\text{T}$: Total memory consumption for the whole simulation (sum for all MPI tasks)
\item $t_\text{W}$: Elapsed wall clock time
\item $t_\text{N}\equiv t_\text{W}\cdot N_\text{N}$: Computational cost in node-hours
\end{itemize}

All benchmarks performed in this work are carried out on the Marconi supercomputer operated by CINECA. The partition Marconi-Fusion available to the European fusion research community comprises of 2396 compute nodes, each equipped with two 24-core Intel Xeon 8160 CPUs (``Skylake'' generation) at a base clock speed of 2.10 GHz and 192 GB DDR4 RAM. The interlink between the compute nodes is provided by Intel OmniPath (100 Gbit/s). We use Intel v.2018.4 compilers with compatible \texttt{intelmpi} and \texttt{mkl} libraries.

\subsection{Outline}\label{:intro:outline}

The rest of the article is organized as follows. Section~\ref{:dev} describes the developments performed as part of the present work. Subsection~\ref{:dev:compl} explains the adaptation to complex preconditioning matrices and subsection~\ref{:dev:modegroups} describes a generalization of the preconditioner, which improves convergence in highly non-linear simulation cases. In Section~\ref{:benchmarks}, previously existing and new solver options are compared based on realistic simulation cases to investigate performance improvements obtained in the present work. Finally, Section~\ref{:summary} briefly provides summary, conclusions, and an outlook to future developments.

\section{Developments highlighted in this work}\label{:dev}

\subsection{Complex representation of preconditioner matrices}\label{:dev:compl}
In order to optimize performance of a direct solver we have exploited the idea of representing the preconditioner block-matrices in complex form. This reduces the memory consumption considerably and improve overall solve time. The transformation into complex format is only possible when the matrix has the appropriate symmetry structure. For example, the following set of equations in the matrix form

\begin{equation}
    \begin{pmatrix}
       a & -b \\
       b & a
    \end{pmatrix}
    \cdot
    \begin{pmatrix}
       x \\
       y
    \end{pmatrix}
    =
    \begin{pmatrix}
       c \\
       d
    \end{pmatrix}
\end{equation}

can be replaced by

\begin{equation}\label{cr2complex}
    \begin{pmatrix}
       a + ib
    \end{pmatrix}
    \cdot
    \begin{pmatrix}
       x + iy
    \end{pmatrix}
    =
    \begin{pmatrix}
       c + id
    \end{pmatrix}
\end{equation}

Since numerical symmetry isn't exactly preserved, the following symmetrization is performed:

\begin{equation}
    \begin{pmatrix}
       a & -b \\
       c & d
    \end{pmatrix}
    \rightarrow
    \begin{pmatrix}
       \left(a+d\right)/2 & -\left(b+c\right)/2 \\
       \left(b+c\right)/2 & \left(a+d\right)/2
    \end{pmatrix}
\end{equation}

This symmetrized matrix can now be replaced by a complex form, analogously to \eqref{cr2complex}. As will be shown later, the approximation introduced by this symmetrization affects the efficiency of the preconditioner only very mildly (small impact on the number of GMRES iterations) such that the overall benefit of using complex matrices prevails. It is to note here that the symmetry enforcement is a crucial step and without it, a convergence can not be achieved. Besides switching to complex input, nothing else was changed in the direct solver library interface. Performance benchmarks of the complex representation of PC matrices are presented in Section~\ref{:realVscmplx:inxflow}.

\subsection{Generalization of the preconditioner for mode groups}\label{:dev:modegroups}
\begin{figure}
\centering
  \includegraphics[height=0.3\textwidth]{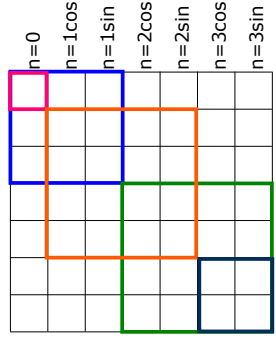}
\caption{Schematic illustration for the \textit{Overlapping} mode group approach. The color blocks outline the parts of the original matrix $A$ used to form the preconditioner matrices which combine modes (0), (0,1), (1,2), (2,3), and (3) into diagonal blocks.}
\label{fig:modegroups}
\end{figure}

In this Section, we describe our approach for improving the efficiency of the preconditioner in highly non-linear scenarios, where the coupling between the toroidal Fourier modes becomes strong causing the GMRES convergence to deteriorate massively or break down entirely. The approach followed here, is to combine several toroidal modes into ``mode groups''. The interaction between the modes within the mode groups is retained in the preconditioning matrix $M$. Consequently, larger block-matrices need to be solved than in the original preconditioner, which makes the solution more expensive in terms of memory consumption and computing time. Nevertheless, capturing the non-linear interaction can improve the convergence of the iterative solver so strongly that the overall performance is significantly enhanced, as demonstrated in Section~\ref{:modegroups:vde}.

While the proposed approach is extremely flexible in terms of choosing mode groups, we illustrate it using the following example of overlapping mode groups (see Figures~\ref{fig:modegroups}). In this example, referred to as \textit{Overlapping}, the preconditioner block-matrices consist of entries from the global matrix $A$ corresponding to a pair of toroidal Fourier modes. Each mode is thus present in two consecutive mode groups. Additional single-mode blocks are added corresponding the first and the last modes.  When the preconditioner solution is formed, each row is filled with the linear combination of the solutions from two corresponding blocks. the weighting factors can be chosen freely. Note, that we don't assemble a complete preconditioner matrix $M$ explicitly, just a preconditioner solution (\ref{eq:PC2}) as needed for the GMRES iterations. In the described approach we effectively take into account interactions of neighboring modes which physically resemble to mode coupling as observed in turbulence cascading, thus we denote this preconditioner as physics based.

Another alternative approach we have considered in this work couples the typically strongest $n=0$ mode with all other $n\ne0$  harmonics by forming $\{(0),\,(0,1),\,(0,2),\dots\}$ mode groups.

The implementation is performed in a very general way allowing more advanced configurations for mode groups than the examples given. It is, for instance, possible to combine modes separated by a certain toroidal period, modes which are believed to have the strongest interactions. Also an automatic and dynamic selection of the groups could be realized in the future. The implementation in JOREK is done as follows:
\begin{itemize}
    \item Via the namelist input file, arbitrary mode groups can be constructed in a flexible way. Each mode group forms a diagonal block in the preconditioner matrix. 
    \item The number of MPI tasks for each mode group can be automatically assigned according to the number of non-zero entries in each mode group, or manually be chosen via the input file. This allows optimizing the load balance for the non-uniform matrix block sizes encountered here.
    \item Mode groups can be overlapping. When reconstructing the whole preconditioner  solution $z=M^{-1}w$ as linear combination of the individual solutions from the mode group blocks, factors are applied. For example, the factor is $1/2$ in the \textit{Overlapping} case. 
\end{itemize}

Mode group methods are benchmarked in Section~\ref{:modegroups:vde} with the standard preconditioner.

\section{Benchmarks for the developments performed}\label{:benchmarks}

Now that we have introduced the developments performed to improve the numerical efficiency of the JOREK solver and preconditioner in the previous section, a set of real-life simulation cases is taken as basis to evaluate the performance of the methods. Each case is briefly explained in the respective subsection. To save computational time, not all solver options are tested for all cases. Note that most developments are already used in production by the JOREK community at the time of submission of this article.

\subsection{Comparison between real and complex solvers in the preconditioner}\label{:realVscmplx:inxflow}
The benchmark is based on a simplified simulation of edge localized modes (ELM) in a JET-like geometry, i.e., an X-point plasma with open and closed field lines included in the simulation domain. A fairly high poloidal grid resolution of 51k Bezier finite elements is used, while only the toroidal mode numbers $n=(0,4,8)$ are included in the simulation. Tests for this case are performed both in the linear phase, where the exponentially growing instabilities are too small to modify the axisymmetric $n=0$ component and in the non-linear phase during the collapse of the edge pressure gradient caused by the instability.

 Table~\ref{tab:303_5_solver_times_real_vs_cmplx} contains data from the simulations restarted in the highly nonlinear regime for a single time-step on 6 and 3 compute nodes with 2 MPI tasks per node. For the given problem size, the real PaStiX~\cite{pastix:1, pastix:2_BLR} solver needs at least 6 compute nodes based on the memory requirement. The complex PaStiX solver, on the other hand, can be used on 3 compute nodes due to the reduced memory consumption.

\begin{table}[ht!]
\begin{center}
\begin{tabular}{ |l|l|c|c|c|} 
 \hline
 \multicolumn{2}{|l|}{PaStiX solver} & real  & \multicolumn{2}{c|}{complex} \\ [0.5ex] 
 \hline
  \multicolumn{2}{|l|}{number of compute Nodes} & 6 & 6 & 3 \\ 
  \hline
 \textbf{A} & Analysis & 22.71 & 14.98 & 11.50 \\ 
 \hline
 \textbf{F} &Factorization & 133.38 & 53.40 & 81.27 \\
 \hline
 \multirow{3}{0pt}{\textbf{I}} & Solve & 1.22 & 0.83 & 1.46 \\
 & Number of GMRES iterations & 25 & 26 & 26\\
 &  GMRES solve & 42.49 & 34.31 & 47.24\\ 
 \hline
\end{tabular}
\caption{Wall-clock time $t_\text{W}$ (sec) spent on solver phases using 6 compute nodes for the real PaStiX solver and 6 resp. 3 compute nodes for the complex PaStiX solver. \label{tab:303_5_solver_times_real_vs_cmplx}}
\end{center}
\end{table}

In Table~\ref{tab:303_5_solver_times_real_vs_cmplx}, we compare the wall-clock time $t_\text{W}$ (sec) spent on solver phases using 6 compute nodes for the real and 6 respectively 3 compute nodes for the complex PaStiX solver.

Due to the reduced matrix dimension, factorization (\textbf{F}) as one of the most expensive parts of the solve step is roughly 2.5 and 3.3 times faster when using the complex solver on 6 and 3 compute nodes, respectively, compared to the real solver. The analysis (\textbf{A}) is also done more efficiently with the complex solver, however, being performed only in the first time step of a simulation, it adds limited benefits for a typical application. Moreover, one can notice from the Table~\ref{tab:303_5_solver_times_real_vs_cmplx} that although the complex solver requires one additional GMRES iteration as compared to the real solver, the GMRES solve (\textbf{I}) on 6 respectively 3 compute nodes is more efficient by a factor of 1.2 and 1.8 as compared to the real solver on 6 compute nodes.

\begin{figure}
\centering
  \includegraphics[width=0.32\textwidth]{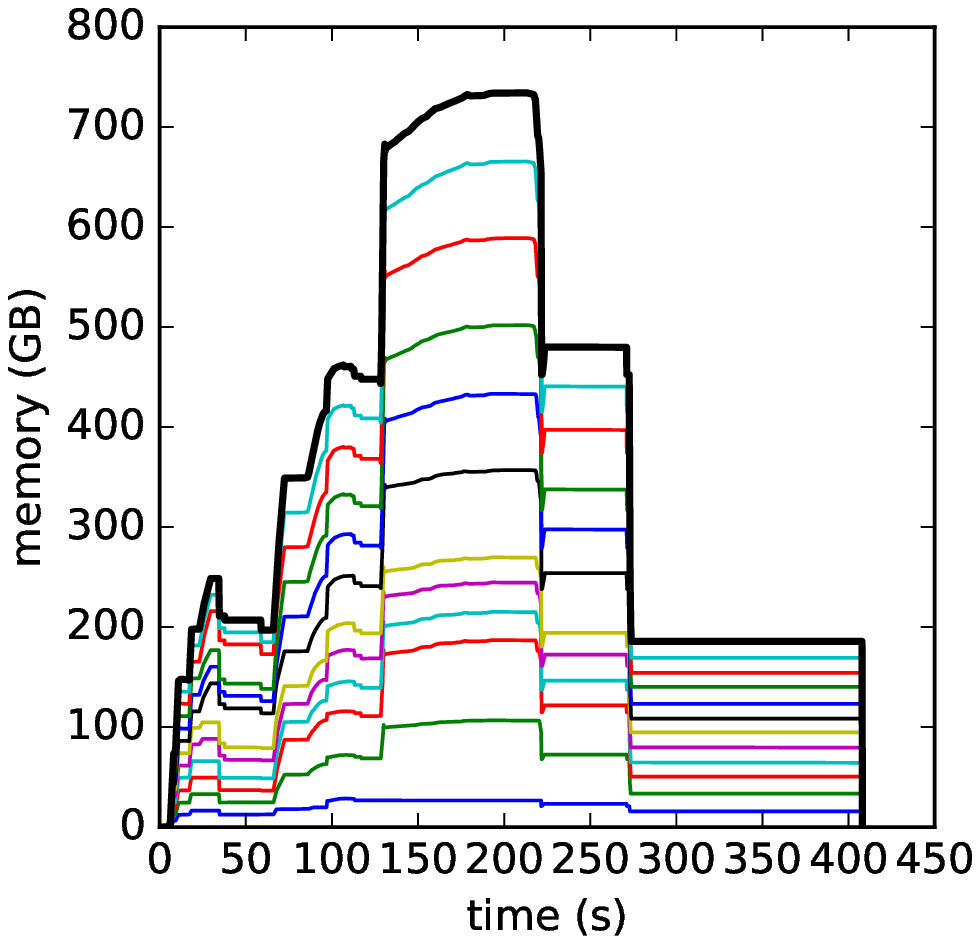}\includegraphics[width=0.32\textwidth]{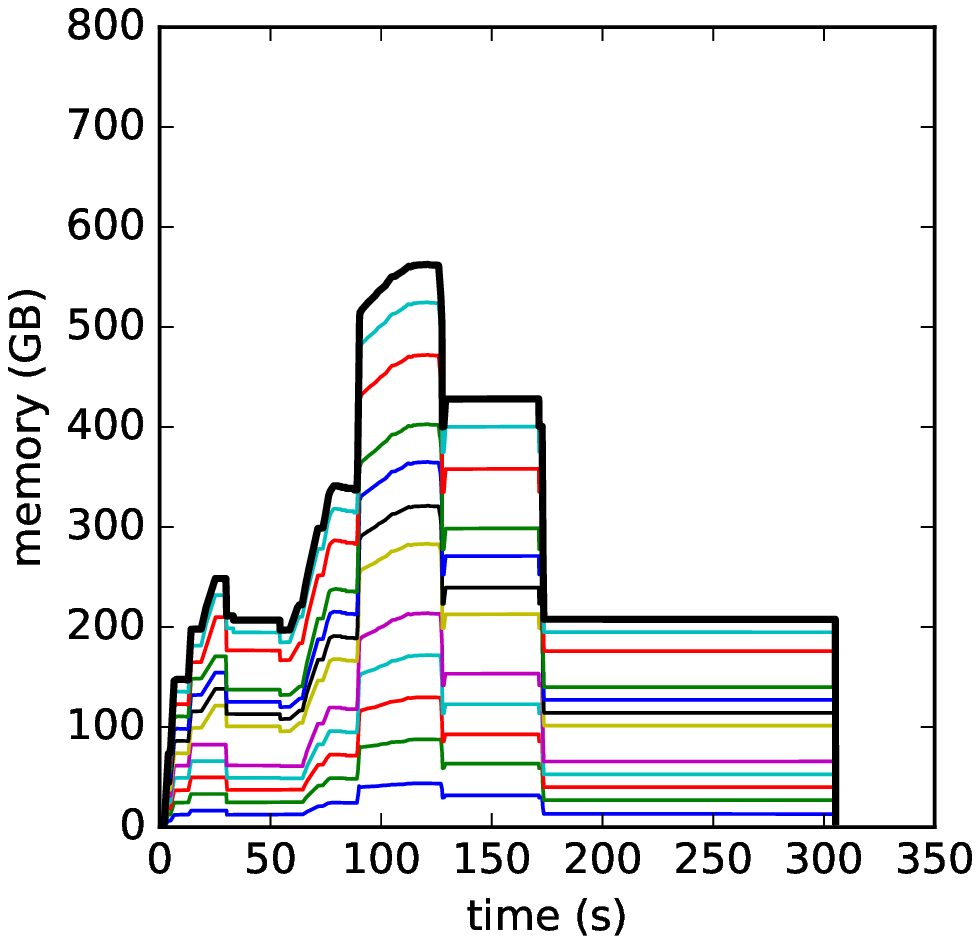}\includegraphics[width=0.32\textwidth]{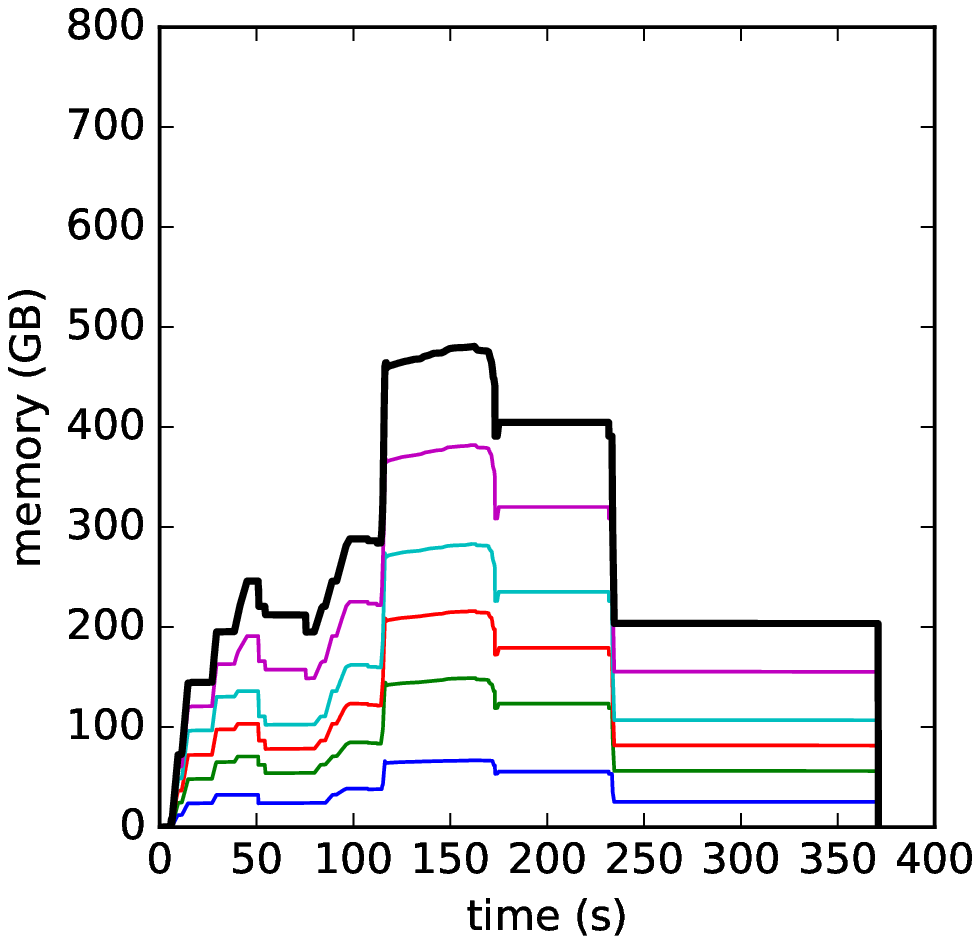}
\caption{Cumulative plot of memory utilization per individual MPI task in the HR simulation case with real (left panel) and complex (middle panel) PaStiX solver using 6 compute nodes, and complex PaStiX solver using 3 compute nodes (right panel)}
\label{fig:8-memory_real_complex}
\end{figure}

Figure~\ref{fig:8-memory_real_complex} shows the time-lines of the memory consumption for the real solver (with 6 compute nodes) and complex solver (with 6 and 3 compute nodes). It is apparent that the maximum memory consumption when using the complex solver with 6 nodes is roughly 600 GB, which is less than roughly 700 GB in the case of the real solver. The memory consumption is even lower in the case on complex solver using 3 compute nodes. Thus, we confirm that the complex solver reduces the overall memory consumption significantly. 
It is worth mentioning that we observe a larger reduction in memory consumption for a higher number of toroidal modes. 

In conclusion, it can be seen from Table~\ref{tab:303_5_solver_times_real_vs_cmplx}, that the overall performance gain by the complex solver is moderate when the memory consumption imposes the same minimum number of compute nodes for the real and complex solvers and when updates of the preconditioning matrix (factorizations) are needed seldomly. On the other hand, if the preconditioning matrix is updated more often, the complex solver may reduce the total computational cost (including matrix construction and preconditioner matrix assembly which are not explicitly discussed in this section) by a third. When the reduced memory consumption by the complex solver allows to use a lower number of compute nodes, the overall computational costs can be approximately halved by moving into a regime where the scalability of the solver improves. 

\subsection{Assessment of computational efficiency with mode groups in the preconditioner}\label{:modegroups:vde}

The simulation considered here is a slightly adapted version of the simulations presented in Ref.~\cite{Artola2020B}. It represents a vertically unstable plasma (vertical displacement event, VDE), that develops violent non-axisymmetric ($n\ne0$) instabilities while moving into the plasma facing components. We have chosen this case, since the standard iterative solver convergence is poor, forcing the use of extremely small time steps. It is suspected that the strong non-linearity of the problem reduces the accuracy by which the preconditioning matrix $M$ approximates the complete system $A$, such that condition number and convergence deteriorate.

We have used this VDE case to test the new preconditioner based on the mode groups. For this purpose we run simulations in the nonlinear regime, and compare the \textit{Standard} preconditioner (Fig.~\ref{fig:modes}) with our new approach based on \textit{Overlapping} mode groups (Fig.~\ref{fig:modegroups}). Another mode-group based approach couples the $n=0$ dominant harmonic with all other modes, referred to as \textit{$n=0$ coupling}.

\begin{table}[ht!]
\begin{center}
\begin{tabular}{ |l|l|c|c|c| } 
 \hline
 & & Standard & Overlapping & $n=0$ coupling \\ [0.5ex] 
 \hline \hline
 \textbf{A} & Analysis time (s) & 27.0 & 99.5 & 56.2 \\ 
 \hline
 \textbf{F} & Factorization time (s) & 21.1 & 123.9 & 58.5 \\
 \hline
 \multirow{2}{0pt}{\textbf{I}} & Solve time (s) & 0.12 & 0.34 & 0.18\\ 
 & GMRES iteration time (s) & 47.3 & 21.3 & 22.6\\
 & Number of GMRES iterations & 130 & 43 & 64\\ 
 \hline
\end{tabular}
\caption{Comparison of solver performance of $n=(0,1,2,3)$ VDE case with $\Delta t=0.05$, based on 10 time steps run with the STRUMPACK library \cite{Ghysels2017}  using 8 compute nodes.\label{tab:vde_preconditioners}}
\end{center}
\end{table}

\begin{table}[ht!]
\begin{center}
\begin{tabular}{ |l|l|c|c|c| } 
 \hline
 & & Standard & Overlapping & $n=0$ coupling \\ [0.5ex] 
 \hline \hline
 \textbf{A} & Analysis time (s) & 24.6 & 126.5 & 56.8 \\ 
 \hline
 \textbf{F} & Factorization time (s) & 22.2 & 147.4 & 60.3 \\
 \hline
 \multirow{2}{0pt}{\textbf{I}} & Solve time (s) & 0.13 & 0.40 & 0.20  \\ 
 & GMRES iteration time (s) & 87.8 & 42.8 & 63.5 \\
 & Number of GMRES iterations & 97 & 45 & 76 \\ 
 \hline
\end{tabular}
\caption{Comparison of solver performance of $n=(0,1,2,\dots,10)$ VDE case with $\Delta t=0.05$, based on 10 time steps run with STRUMPACK library using 22 compute nodes.\label{tab:vde_preconditioners_n11}}
\end{center}
\end{table}

As shown in Table \ref{tab:vde_preconditioners}, the \textit{Overlapping} preconditioner reduces the number of GMRES iterations by a factor of 3, leading to the reduction of the whole GMRES cycle time by a factor of 2.2, respectively. Meanwhile, the factorization time increased by a factor of 6. Thus, the overall performance depends on how often preconditioner updates are required. A similar improvement trend with the \textit{Overlapping} method is observed for the VDE case with higher toroidal resolution including $n=(0,1,2,\dots,10)$, as shown in Table~\ref{tab:vde_preconditioners_n11}.

The analysis, factorization and solve times with the new preconditioner are determined by the corresponding times spent on the largest diagonal block, which is twice the size of the standard case. Since each diagonal block-matrix is factorized and solved independently, the difference in matrix sizes introduces a strong workload imbalance. To address this, we have implemented a flexible distribution of MPI tasks to each mode group. For example, in the current simulation using the \textit{Overlapping} method, we have five mode groups and a total of 32 MPI tasks. The groups consist of the following toroidal modes: $\{(0),(0,1),(1,2),(2,3),(3)\}$. The block size corresponding to $n=0$ mode is twice smaller than for any other mode, and the number of nonzero entries is 4 times smaller. The preconditioner block-matrix sizes for each group are shown in Table \ref{tab:vde_families}.

\begin{table}[h!]
\begin{center}
\begin{tabular}{ |l|c|c|c|c|c| } 
 \hline
 Mode sets & (0) & (0,1) & (1,2) & (2,3) & (3) \\ [0.5ex] 
 \hline \hline
 Matrix rank & 200,767 & 602,301 & 803,068 & 803,068 & 401,534 \\ 
 nnz & 52,808,329 & 475,274,961 & 844,933,264 & 844,933,264 & 211,233,316 \\
 \hline
\end{tabular}
\caption{Ranks and numbers of nonzero entries in preconditioner block-matrices for $n=(0,1,2,3)$ VDE case using the \textit{Overlapping} preconditioning method.\label{tab:vde_families}}
\end{center}
\end{table}

There are different ways to distribute the available 32 MPI tasks among the mode groups. For instance, distributing them as (2,6,9,9,6) would increase the factorization time to 147.6 sec as compared to 123.9 sec in the case of (1,4,12,12,3) tasks per group distribution. Here one should take into account that the most computationally expensive factorization has a different MPI task scaling compared to the GMRES iteration cycle, thus the optimal task distribution for balanced performance depends on the relative contribution of total factorization time and total iteration time. However, as a rule of thumb, one can distribute tasks roughly proportionally to the number of nonzero entries in each mode group. In $n=(0,1,2,3)$ we have assigned 1 and 3 MPI tasks to the single mode groups of the $(0)$ and the $(3)$ modes, respectively, 4 tasks to the $(0,1)$ group, and 12 tasks to each of the $(1,2)$ and the $(2,3)$ groups. In the $n=(0,1,2,\dots,10)$ case 2 MPI tasks are assigned to the smallest mode group of $(0)$ mode, 6 MPI tasks are assigned to the mode group of $(11)$ mode, and the overlapping mode groups get 8 MPI tasks each.

Since the preconditioning matrix needs not to be updated (factorized) in every time step, the efficiency of the new preconditioner is determined by the actual frequency of these updates. In our example, the \textit{Overlapping} preconditioner would lead to the same overall performance, if the preconditioner is updated every five time steps. This number would be slightly higher in practice, since the time to construct preconditioner matrices is also slightly higher for the new preconditioner. In the nonlinear phase of the VDE simulation with a normalized time step of $0.05$, the preconditioner matrix updates need to be performed approximately every 50 time steps, which gives the new preconditioner a strong advantage.

Meanwhile, as the preconditioner approximates the total system a lot better now, it becomes possible to increase the simulation time step allowing for an even more efficient way of running the simulation. For the same nonlinear case, good overall performance was obtained when increasing the time step by a factor of $3$ for the \textit{Overlapping}  preconditioner, leading to the total $t_\text{W}$ spent on simulating 100 steps using 8 compute nodes of 104 min. Comparing this to 97 min required to perform 100 steps with $\Delta t=0.05$ using the \textit{Standard} preconditioner, an approximate speed-up of a factor $3\times$ is obtained. For the \textit{$n=0$ coupling} preconditioner the time step could be increased by a factor of $2$ leading to the total simulation time of 95 min, i.e. a factor $2\times$ speed-up compared to the \textit{Standard} approach. The same conclusion holds for the VDE case with a higher toroidal resolution including the $n=(0,\dots,10)$ toroidal modes, which was tested as well. 

\begin{figure}
\centering
  \includegraphics[width=0.45\textwidth]{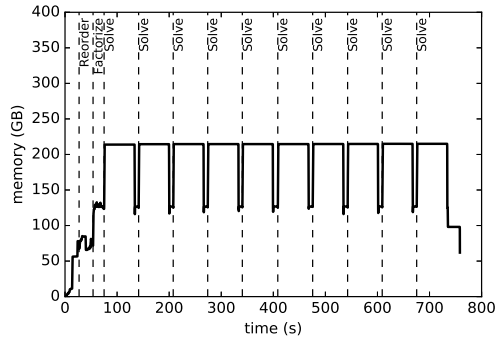} 
\caption{Cumulative memory consumption in the $n=(0,1,2,3)$ VDE case simulation with \textit{Standard} preconditioner using 8 compute nodes with 4 MPI tasks per node}
\label{fig:vde7_memory_standard}
\end{figure}

\begin{figure}
\centering
  \includegraphics[width=0.45\textwidth]{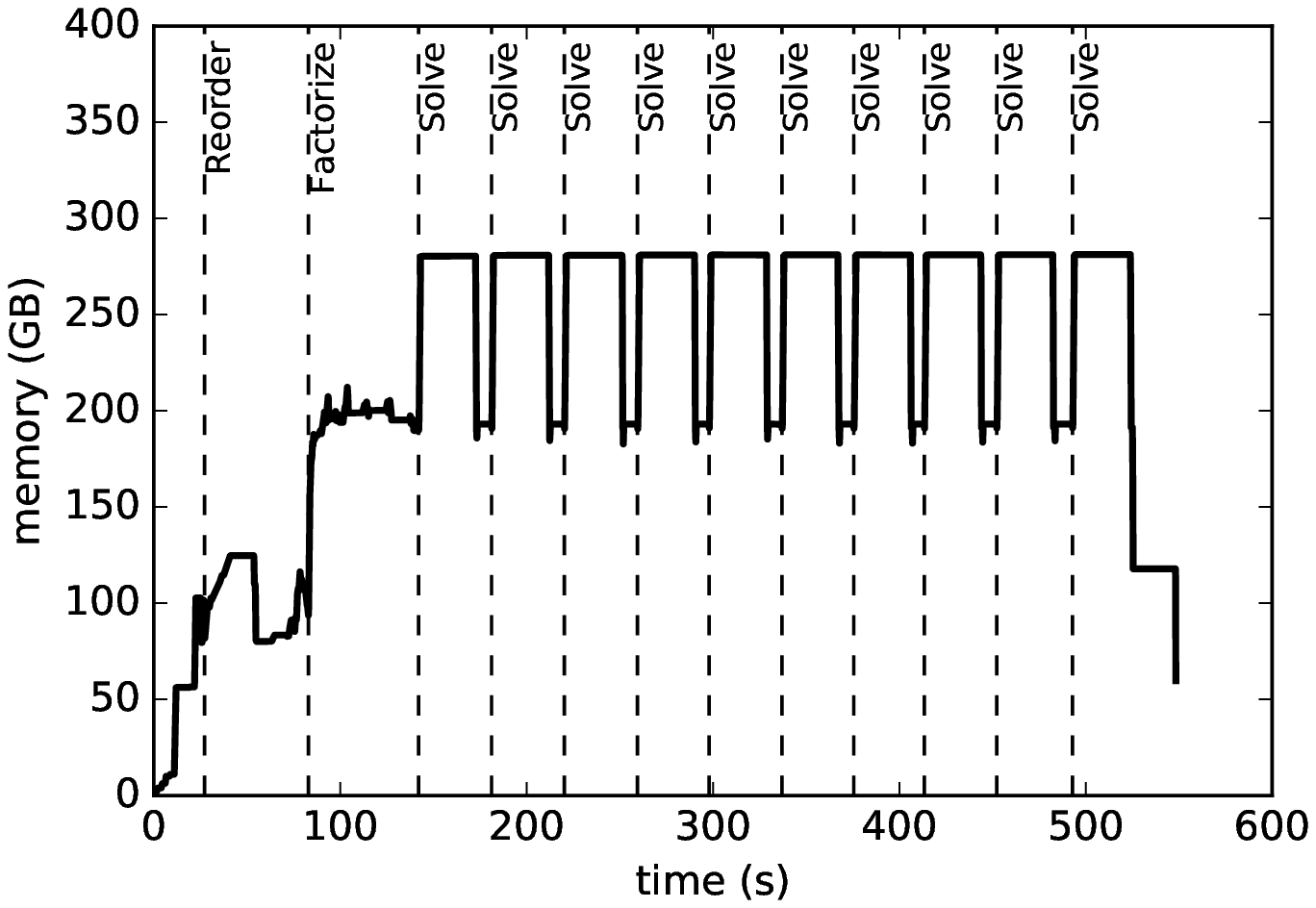} \includegraphics[width=0.45\textwidth]{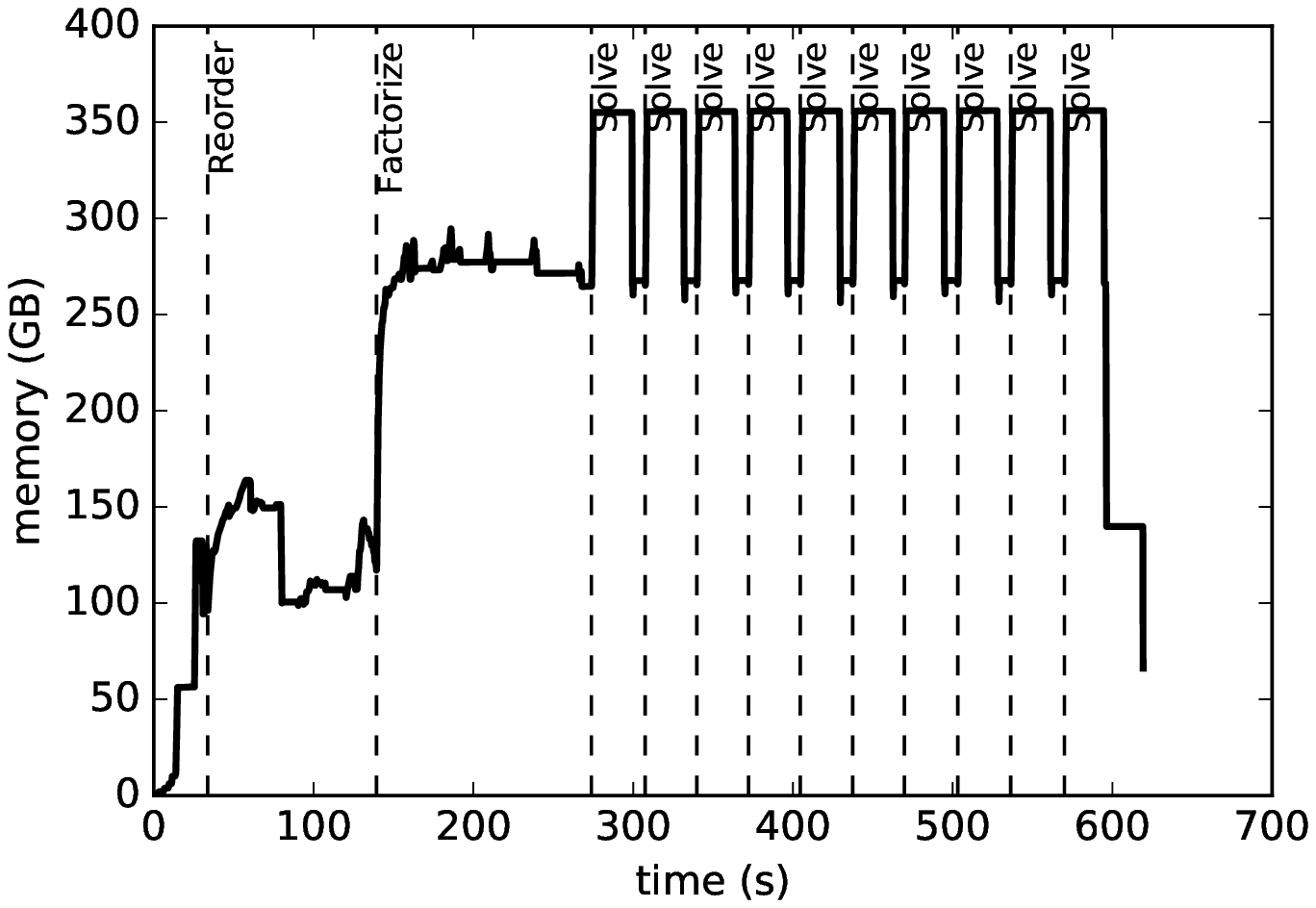}
\caption{Cumulative memory consumption in the $n=(0,1,2,3)$ VDE case simulation with \textit{$n=0$ coupling} preconditioner (left panel) and \textit{Overlapping} preconditioner (right panel) using 8 compute nodes with 4 MPI tasks per node}
\label{fig:vde7_memory_new}
\end{figure}

The main disadvantage of the new preconditioner lies in the increased memory consumption. As we can see in the Figures~\ref{fig:vde7_memory_standard}--\ref{fig:vde7_memory_new}, for the VDE case with $n=(0,1,2,3)$, the maximum required memory with the \textit{Overlapping} preconditioner is $\sim 75\%$ higher than for the \textit{Standard} method, while for the \textit{$n=0$ coupling} preconditioner the overall memory consumption increases by $\sim 30\%$.

The presented results demonstrate just an example of using the new preconditioner. With the provided implementation flexibility, there is room for further performance improvements, e.g. by adjusting work load balance, data locality, optimization of GMRES preconditioner update frequency vs.\ simulation time step, etc.


\section{Summary}\label{:summary}

In this article, we have described the iterative solver with the physics-based preconditioner used in the non-linear MHD code JOREK, the properties of the sparse matrix system, and the interfaces to parallel sparse matrix solver libraries. We also explained some limitations faced with this solver in particular regarding memory consumption and convergence in highly non-linear scenarios.

The new interface was developed, which allows using complex numbers in the preconditioner, leading to improved performance and memory utilization. Significantly reduced memory consumption with the complex preconditioner allows to perform a simulation with less compute nodes and thus allows to shift the point of operation towards the lower (ideal) part of the strong scaling curve. 

The most important development is a generalization of the preconditioner to ``mode groups''. The new method allows capturing some of the non-linear mode coupling in the PC by combining several toroidal modes in the diagonal matrix blocks, which can be solved independently. This preconditioner resembles the original system significantly better in nonlinear scenarios, thus leading to major improvement of the GMRES convergence and, consequently, to a significant reduction of the overall run time. Besides the benefits for tokamak simulations, the mode groups will also be an essential building block to make future JOREK simulations of stellarator cases \cite{Nikulsin2019,Ramasamy2020} possible, where instabilities do not linearly decouple in the toroidal modes any more. the developments can be combined with additional (radial) parallelization strategies in the future.

\section*{Acknowledgements}

The authors would like to thank Omar Maj for fruitful discussions, and FJ Artola, A Cathey, F Wieschollek, I Krebs, R Ramasamy for providing ``real-life'' test cases, Mathieu Faverge and the rest of the PaStiX team for their assistance with PaStiX 6. One of the author (I.H.) would like to thank Sebastian Ohlmann (MPCDF) and Pieter Ghysels (LBNL) for technical advising.

Part of this work has been carried out within the framework of the EUROfusion Consortium and has received funding from the Euratom research and training program 2014-2018 and 2019-2020 under grant agreement No 633053. The views and opinions expressed herein do not necessarily reflect those of the European Commission. Some of the work was performed using the Marconi-Fusion supercomputer.

{\footnotesize
\bibliography{main}
}

\end{document}